\documentclass{iaus260}
\usepackage{graphicx}
\pubyear{2009}
\volume{260}  
\pagerange{10--17}
\jname{The R\^ole of Astronomy in Society and Culture}
\editors{D. Valls-Gabaud \& A. Boksenberg, eds.}

\title[Aboriginal Astronomy]                   
{The Astronomy of Aboriginal Australia}  

\author[Norris \& Hamacher]           
{Ray P. Norris$^{1,2}$ \and Duane W. Hamacher$^2$}        

\affiliation{$^1$CSIRO Australia Telescope National Facility\\ 
                 PO Box 76, Epping, NSW, 1710, Australia\\ 
                 email: {\tt Ray.Norris@csiro.au} \\[\affilskip]
             $^2$Department of Indigenous Studies, Macquarie University\\ 
                 North Ryde, NSW 2109, Australia \\
                 email: {\tt Duane.Hamacher@mq.edu.au}}

\begin{document}
\maketitle

\begin{abstract}
The traditional cultures of Aboriginal Australians include a significant astronomical component, which is usually reported in terms of songs or stories associated with stars and constellations. Here we argue that the astronomical components extend further, and include a search for meaning in the sky, beyond simply mirroring the earth-bound understanding. In particular, we have found that traditional Aboriginal cultures include a deep understanding of the motion of objects in the sky, and that this knowledge was used for practical purposes such as constructing calendars. We also present evidence that traditional Aboriginal Australians made careful records and measurements of cyclical phenomena, and paid careful attention to unexpected phenomena such as eclipses and meteorite impacts. 

\keywords{Australia, Aboriginal, archaeoastronomy, ethnoastronomy, history of meteoritics.}    
\end{abstract}

\firstsection 

\section{Introduction}

Before the British occupation of Australia, there were an estimated 400 different Aboriginal cultures. Each had its own language, stories, and beliefs, although most were centred on the idea that the world was created in the ``Dreaming" by ancestral spirits, whose presence can still be seen both on the land and in the sky.  These ancestral spirits taught humans how to live, and left a user guide to life in their songs and stories. Many of these stories are reflected in the patterns in the sky. This is reflected in existing literature on Australian Aboriginal Astronomy, which tends to focus on stories associated with constellations, but with little emphasis on the deep intellectual context in which meaning is sought for astronomical phenomena.

The 50,000 year-old Aboriginal cultures are believed to be the oldest continuous cultures in the world, and since the night sky seems to play an important role in them, it is sometimes suggested (e.g.  \cite[Haynes 1992]{Haynes}) that Aboriginal Australians were the world's first astronomers.  But the word ``astronomy" implies more than just stories -- it means a quest to understand the patterns in the sky, and the motion and eclipses of the Sun, Moon, and planets.  The goal of our research project is to understand the importance of Astronomy in Aboriginal Cultures, and explore the depth of traditional Aboriginal astronomical knowledge.

The project has two key components. In some parts of Australia, such as Arnhem Land in the north of Australia, Aboriginal cultures are flourishing. For example, Yolngu people still maintain traditional aspects of their lifestyle, and continue to conduct initiation ceremonies, in which traditional knowledge is passed from generation to generation. The first thread of the project aims to record their stories and ceremonies, and as much of the astronomical lore as might be told to an uninitiated white person. 
  
   In other parts of Australia, the Aboriginal culture was badly damaged by the arrival of Europeans some 200 years ago. For example, the Aboriginal people around Sydney were decimated by introduced disease, exclusion from food and water, and deliberate genocide. In these regions, little is known of the original culture of the Aboriginal people, but we can study it by examining their art and artefacts. Thus the second thread of the project focuses on surveying and recording the rock engravings of the Sydney basin region and the stone arrangements of Victoria.  
   
   Aboriginal astronomy was first described by \cite[Stanbridge (1857)]{Stanbridge}, and subsequent important works include those by \cite[Mountford (1976)]{Mountfordb}, \cite[Haynes (1992)]{Haynes}, \cite[Johnson (1998)]{Johnson}, and \cite[Cairns \& Harney (2003)]{CairnsHarney}. Most of these works focus on how objects in the night sky represent events or characters in Dreaming stories, and only touch briefly on practical applications or on interpretation of the motion of the sky.
   
   For example, the European constellation of Orion is called Djulpan in the Yolngu language. The three stars in Orion's belt are three brothers who were blown into the sky by the Sun after one of them broke a taboo by eating a king-fish, which was his totem animal \cite[Wells (1973)]{Wellsb}. They are seated in a canoe, with Betelgeuse marking the front of the canoe, and Rigel the back of the canoe. The Orion nebula is the fish, and the stars of Orion's sword are the line still attached to the fish (Fig.\,\ref{emu} - Left).  Different stories about Orion are found in other Aboriginal cultures, but are often associated with young men, particularly those who are hunting or fishing (e.g. \cite[Massola 1968]{Massola}).

Similarly, many groups across Australia (e.g. \cite[Massola 1968]{Massola}, \cite[Harney 1959]{Harney}) tell how the Pleiades are a group of sisters chased by the young men in Orion, which is very similar to the traditional European myth about these constellations. Although this similarity between Aboriginal and European stories suggests early cultural contact between Aboriginal and European people, it is unlikely that such contact took place. It is more likely that the Aboriginal people independently devised the stories in a sort of cultural convergent evolution, perhaps reflecting the fact that the massive bright stars of Orion follow the beautiful little cluster of the Pleiades as the sky rotates.

\section{Cultural Background}
  
   An impediment to this study is the misinformation permeating the literature, dating from an era when racial stereotypes were widespread even amongst academics. For example, \cite[Blake (1981)]{Blake} states categorically ``No Australian Aboriginal language has a word for a number higher than four'' and this belief is still widely encountered. However, complex Aboriginal number systems have been well-documented in the literature (e.g. \cite[McRoberts 1990]{McRoberts}, \cite[Tully 1997]{Tully}), and Blake's statement appears to reflect a widely-held post-colonial attitude to Aboriginal people which was widespread fifty years ago but is now fortunately waning. However, such ingrained attitudes state equally misleadingly that Aboriginal people ``don't measure things'', and so would not be interested in or capable of careful astronomical measurements. Rather than relying on such assertions, this project concentrates on exploring the available evidence.

\section{Sun, Moon, and Eclipses }

   In most Aboriginal cultures, the Moon is male and the Sun is female. For example, the Yolngu people of Arnhem Land in the far north of Australia, tell how Walu, the Sun-woman, lights a small fire each morning, which we see as the dawn (\cite[Wells 1964]{Wellsa}). She decorates herself with red ochre, some of which spills onto the clouds, creating the red sunrise. She then lights her torch, made from a stringy-bark tree, and travels across the sky from east to west carrying her blazing torch, creating daylight. On reaching the western horizon, she puts out her torch, and starts the long journey underground back to the morning camp in the east. Thus the Yolngu people explained the daily motion of the Sun across the sky and back again under the ground. 

This raises the question of whether traditional Aboriginal people regarded the Earth as being flat, in which case the Moon would need to tunnel under the Earth, or whether they regarded the Earth as being a finite body which the Sun might pass underneath. \cite[Warner (1937)]{Warner} was told by a traditional Yolngu man that ``the Sun goes clear around the world", and who illustrated this ``by putting his hand over a box and under it and around again".
   
   The Yolngu people call the Moon Ngalindi and he too travels across the sky. Originally, he was a fat lazy man (corresponding to the full Moon) for which he was punished by his wives, who chopped bits off him with their axes, producing  the waning Moon (\cite[Wells 1964]{Wellsa}, \cite[Hulley 1996]{Hulley}). He managed to escape by climbing a tall tree to follow the Sun, but was mortally wounded, and died (the new Moon). After remaining dead for three days, he rose again, growing round and fat (the waxing Moon), until, after two weeks his wives attacked him again. The cycle continues to repeat every month. Until Ngalindi first died, everyone on Earth was immortal, but he cursed humans and animals so that only he could return to life. For everyone else, death would thereafter be final. 
   
   But the Arnhem Land stories go much further, even explaining why the Moon is associated with tides, and in particular why Spring tides are associated with the Full Moon and the New Moon. When the tides are high, water fills the Moon as it rises. As the water runs out of the Moon, the tides fall, leaving the Moon empty for three days. Then the tide rises once more, refilling the Moon.  So, although the mechanics are a little unclear, the Yolngu people clearly had an excellent understanding of the motions of the Moon, and its relationship to the tides.
   
   The Warlpiri people explain a solar eclipse as being the Sun-woman being hidden by the Moon-man as he makes love to her. On the other hand, a lunar eclipse is caused when the Moon-man is chased by the Sun-woman who is pursuing him, and eventually catches-up.  These two stories demonstrate an understanding that eclipses were caused by the paths of the Sun and Moon across the sky, occasionally intersecting (\cite[Johnson 1998]{Johnson}, \cite[Warner 1937]{Warner}). This realisation is found in several other language groups. For example, \cite[Bates (1944)]{Bates} recounted how, during the solar eclipse of 1922, the Wirangu people told her that the eclipse was caused by the Sun and Moon ``becoming husband and wife together''.

These stories of eclipses are remarkable for two reasons. First, solar eclipses are rare, and a total solar eclipse is seen in any one location only once every three or four generations. So for any group to have an explanation of solar eclipses implies a remarkable continuity of learning. For this myth to have been created, someone must have seen an eclipse, related it to a story passed to them from several generations previously, and then strengthened and retold it for someone a few generations later.

Secondly, to relate a lunar eclipse to the paths of the Sun and Moon implies a remarkable intellectual feat of visualisation of their motion. At a lunar eclipse, the Sun, Earth, and Moon are in a line, with the Earth between the Sun and the Moon. It is an impressive intellectual feat, given that the Sun and Moon are diametrically opposed, to reason that it is precisely this alignment which has caused the eclipse.

   \begin{figure}[h]
\begin{center}
 \includegraphics[width=1.4in]{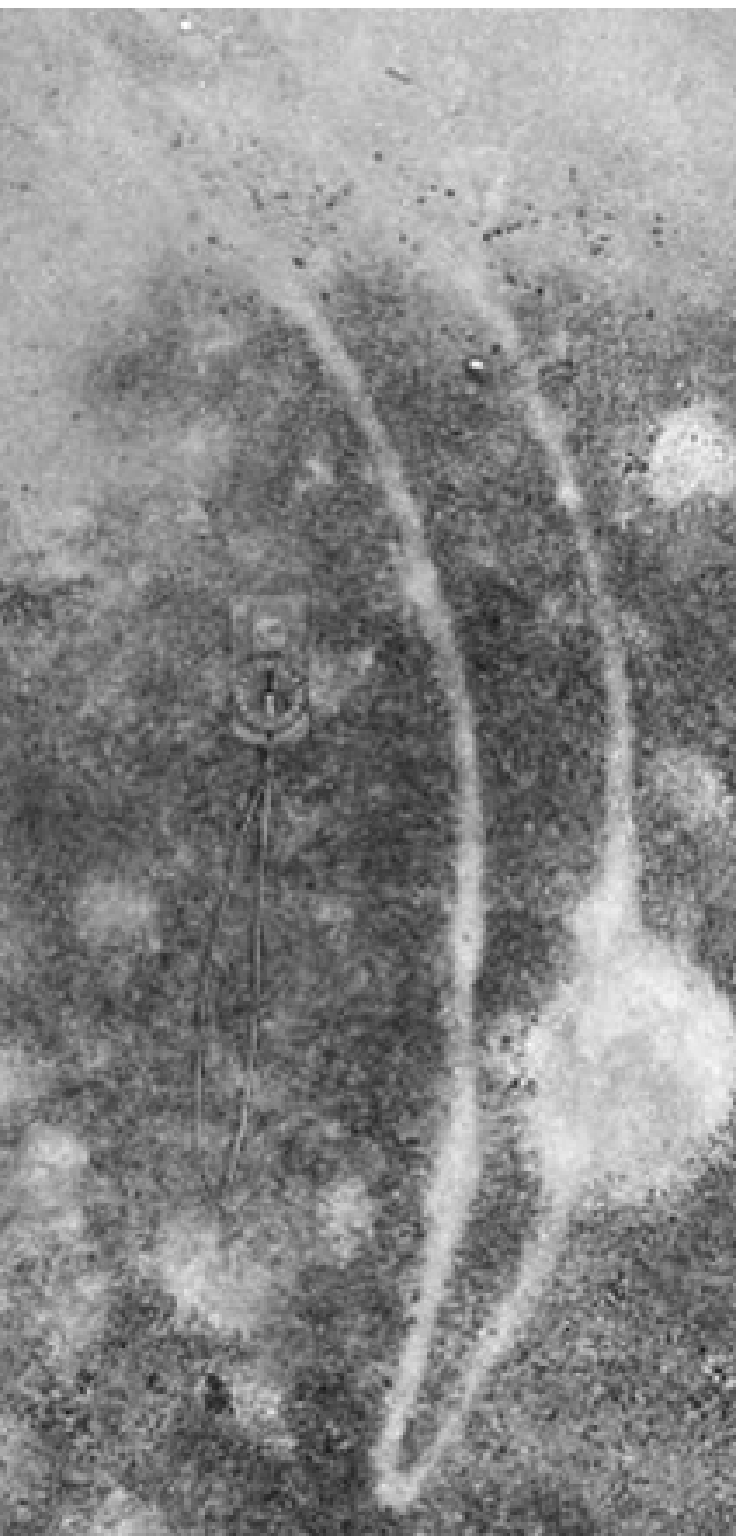} 
  \includegraphics[width=1.4in]{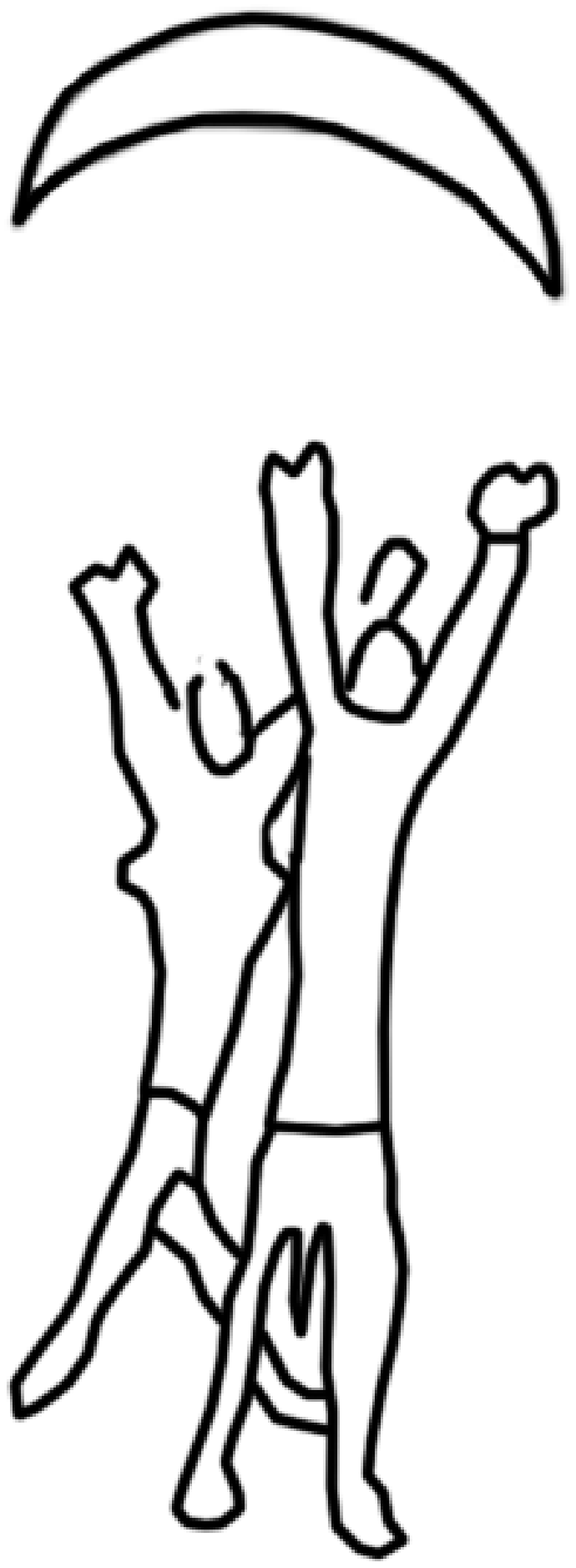} 
 \caption{(Left) A rock engraving showing a crescent.  (Right) A rock engraving of a man and woman reaching up to a crescent.  Could this represent an eclipse?}
  \label{moon}
\end{center}
\end{figure}
   
   Amongst thousands of beautiful rock engravings in Ku-ring-gai Chase National Park, just north of Sydney, are a number of crescent shapes, such as that shown in Fig.\,\ref{moon} (Left). Archaeologists (e.g. \cite[McCarthy 1983]{McCarthy}) have  traditionally referred to these shapes as boomerangs. However, a detailed study (\cite[Norris \& Hamacher 2009]{Norrisb}) has shown that these shapes are more likely to represent crescent moons than boomerangs. For example, boomerangs usually have two straight lengths rather than a single curved crescent, and rarely have pointed ends. Furthermore, it is unclear why a man and woman should reach up towards a boomerang in the sky. But if these shapes are moons, then why is the moon shown with the two horns pointing down, since that configuration is seen only in the afternoon or morning when the Sun is already high in the sky, and the moon barely visible?
   
   One answer is that it might depict an eclipse. In Fig.\,\ref{moon} (Right), the man stands in front of the woman, partly obscuring her. Such carefully-drawn obscurations are unusual in these rock carvings, and in this case may well represent the Moon-man obscuring the Sun-woman during a solar eclipse.

\section{The Calendar}

   Aboriginal calendars tend to be more complex than European calendars, and those in the north of Australia are often based on six seasons. Some Aboriginal groups mark them in terms of the stars which appear during these seasons. For example, the Pitjantjatjara people say that the rising of the Pleiades in the dawn sky in May heralds the start of winter (\cite[Clarke 2003]{Clarke}).   Perhaps even more importantly, the heliacal rising of a star or constellation can tell people when it's time to move to a new food source. For example, when the Mallee-fowl constellation (Lyra) appears in March, the Boorong people of Victoria know that the Mallee-fowl are about to build their nests, and when Lyra disappears in October, the eggs are laid and are ready to be collected (\cite[Stanbridge 1857]{Stanbridge}). Similarly, the appearance of Scorpius told Yolngu people that the Macassan (Indonesian) fisherman would soon arrive to fish for Trepang. 
   
\begin{figure}[h]
\begin{center}
 \includegraphics[width=2.5in]{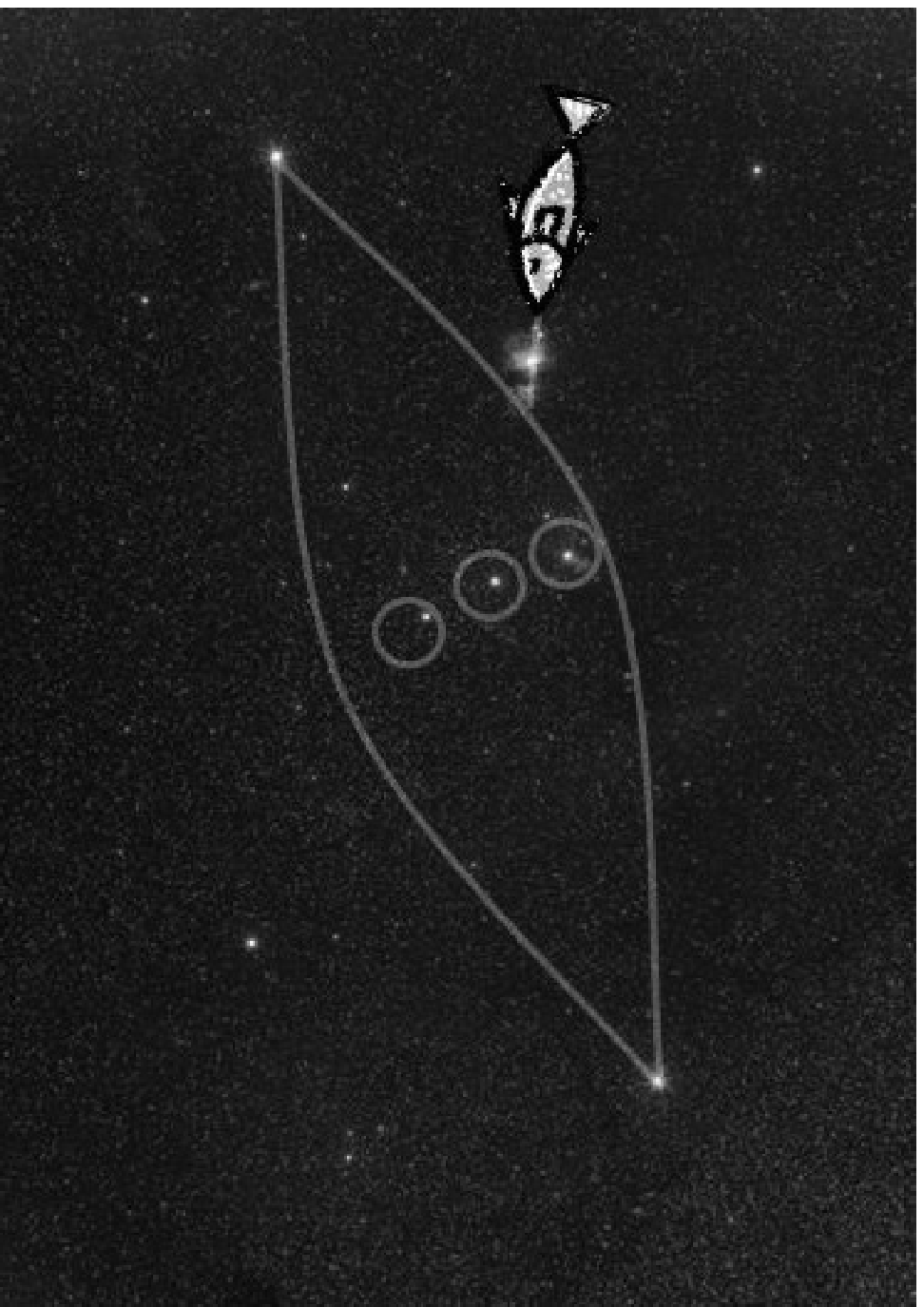}
 \includegraphics[width=2.36in]{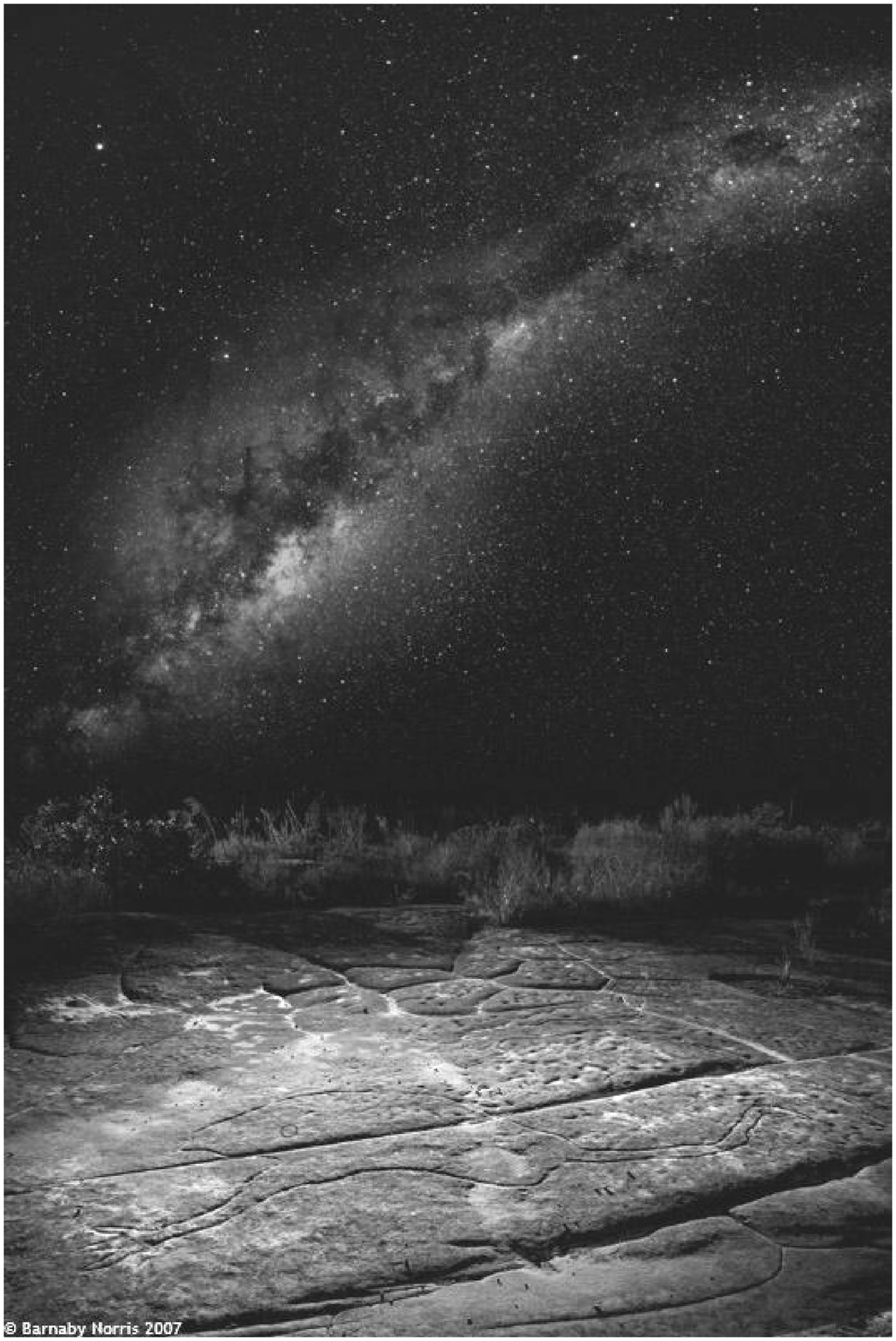} 
 \caption{(Left) The canoe, Djulpan, in Orion.  (Right) The emu in the sky above her engraving.  \copyright Barnaby Norris, 2007}
   \label{emu}
\end{center}
\end{figure}
   
   Close to the Southern Cross (a possum in a tree, according to the Boorong people) is a dark cloud of interstellar dust, called the Coalsack by astronomers.  To many groups, it's the head of the ``Emu in the Sky".  The emu's body stretches down to the left towards Scorpius, dominating the southern Milky Way. In Ku-ring-gai Chase National Park is an engraving of an emu, which appears to be oriented (\cite[Cairns 1996]{Cairns}) to line up with the Emu in the Sky, in the correct orientation, at just the time of year when real-life emus are laying their eggs (Fig.\,\ref{emu} - Right). 

\section{Morning and Evening Star}

   When the planet Venus appears in the sky as a Morning Star, Yolngu people call her ``Banumbirr'', and tell how she came across the sea from the east in the Dreaming, naming and creating animals and lands as she crossed the shoreline, and continued travelling westwards across the country, leaving as her legacy one of the ``songlines'' which are important in Aboriginal cultures, and to Aboriginal navigation.
   
   In an important ``Morning Star Ceremony'', earthly Yolngu people communicate with their ancestors living on Baralku, the island of the dead, with the help of Banumbirr together with a ``Morning Star Pole''. The ceremony starts at dusk and continues through the night, reaching a climax when Banumbirr rises a few hours before dawn. She is said to trail a faint rope behind her along which messages are sent, and which prevents her from ever moving away from the Sun. This faint line in the sky is probably zodiacal light, which is caused by extraterrestrial dust in the plane of the solar system.
   
   The Morning Star ceremony tells us two important things. One is that Yolngu people observed that Venus never strays far from the Sun, which they explain in terms of the rope binding Venus to the island. The other is that the Morning-Star ceremony has to be planned well in advance, since Venus rises before dawn only at certain times of the year, which vary from year to year. So the Yolngu people also track the complex motion of Venus well enough to predict when to hold the Morning Star Ceremony.

\cite[Wells (1996, 1973)]{Wellsa, Wellsb} tells a corresponding story about Djurrpun, the Evening Star, which she assumes also to be Venus.  She relates how the  appearance of the Evening Star shortly after sunset tells the Yolngu people that it's time to start gathering ``rakai'' which are the corms of the lotus lily, and an important food-source for Yolngu people. However, from an astronomical perspective this story must be faulty, since the appearance of Venus as an evening star does not occur at a particular time of year, but varies across the seasons, rendering it useless for this purpose. This puzzle was solved when one of us (\cite[Norris 2008]{Norris}) was told the story by the traditional custodian of this story, Mathulu Munyarryun. The Evening Star in this context is not Venus, but the star Spica, which does indeed set just after the Sun at the time of year when rakai nuts are ready for harvesting.
   
\section{Comets, Meteors and Cosmic Impacts}

Amongst the Dreaming stories of various Aboriginal groups, there are many different views of comets and meteors.    In a universe that seems ordered and predictable, transient events such as comets and meteors were often seen with fear and apprehension.  As a result, comets and meteors are frequently associated with omens, death, and evil.  The Tiwi of Bathurst and Melville Islands (\cite[Mountford 1958]{Mountford}) and the Kuninjku of Arnhem Land (\cite[Taylor 1996]{Taylor}) saw meteors as the fiery eyes of evil spirit beings who raced across the sky, hunting for the souls of the sick and dying.  When children from the Ooldea Region of South Australia saw a meteor (which they called a ``devil-devil"), they would chant ``\textit{Kandanga daruarungu manangga gilbanga}" which means ``\textit{star falling at night time go back!}" (\cite[Harney \& Elkin 1949]{HarneyElkin}).  The Ngarrindjeri of South Australia told of a being named Kulda who would manifest as a meteor emerging from the Southern Cross, warning the people of a disease epidemic.  This led the people to shout, ``\textit{peika baki}", meaning ``\textit{death is coming}" (\cite[Tindale 1933/34]{Tindale}).  The arrival of the Great Comet of 1843 caused fear among the Ngarrindjeri and was seen as a harbinger of calamity (\cite[Eyre 1845]{Eyre}).  These views are shared by peoples throughout the world (e.g. \cite[Frazer 1930]{Frazer}, \cite[Hamacher \& Norris 2009]{Hamacher}). 

Occasionally, one of these falling stars actually crashes to earth.  There are many Aboriginal stories that describe such an event, including the Arrernte story about the creation of Tnorala, or Gosse's Bluff, an impact crater created nearly 150 million years ago.  In the Dreaming, a group of sky-women took the form of stars and danced in the Milky Way.  One of the women grew tired and placed her baby in a wooden basket, called a turna.  As the women continued dancing, the turna fell over the edge and plunged to earth.  The baby fell and was covered by the turna.  The impact forced the rocks upward, forming the walls of the circular structure.  The baby's mother, the evening star, and father, the morning star, continue to search for their baby to this day (\cite[Cauchi 2003]{Cauchi}, \cite[Williams 2004]{Williams}).  A Yolngu story from Arnhem Land tells of Goorda, a lonely fire spirit that lived in the Southern Cross, who came to earth as a meteor to bring fire to the people.  Upon touching the ground near the Gainmaui River, he accidentally caused massive fires that brought death and destruction to the people (\cite[Allen 1975]{Allen}).  Various Dreaming stories from across Australia tell about falling stars that bring fire to the earth (e.g. \cite[Peck 1933]{Peck}, \cite[Berndt \& Berndt 1988]{Berndt}, \cite[Jones 1989]{Jones}).  Perhaps some of these stories record meteorite falls and cosmic impacts that remain unknown to Western science?  Hopefully, further research will tell.

\section{Astronomical Measurements}
   Having established that traditional Aboriginal cultures embodied a deep interest in the motion of heavenly bodies, can we find any evidence to support the ``Stonehenge hypothesis'' that careful observations were made, records kept, or structures set up to point to the rising and setting places of heavenly bodies?
   
   
   
\begin{figure}[h]
\begin{center}
 \includegraphics[width=4in]{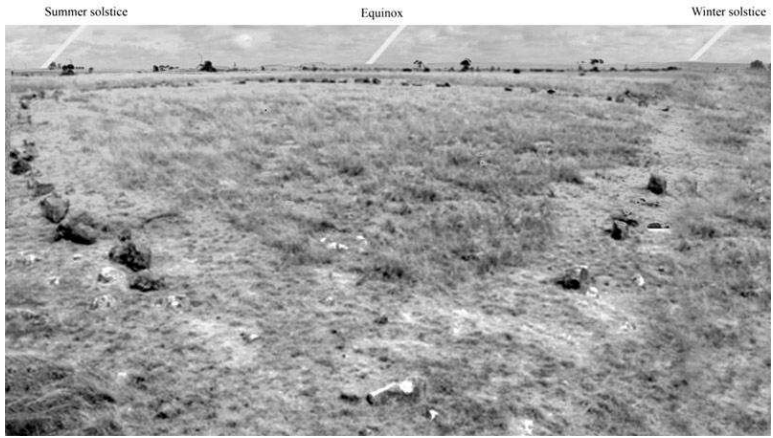} 
 \caption{The view across the Wurdi Youang stone circle, showing the positions of the setting sun at the solstices and equinox.  Lower part of the composite \copyright John Morieson.}
   \label{wurdi}
\end{center}
\end{figure}

   A prime example is the Wurdi Youang stone arrangement in Victoria, which was built by the Wathaurung people before European settlement. This egg-shaped ring of stones, about 50m in diameter, has its major axis almost exactly East-West. At its Western end, at the highest point of the circle, are three prominent waist-high stones.  \cite[Morieson (2003)]{Morieson} pointed out that some outlying stones to the West of the circle, as viewed from these three stones, seem to indicate the setting positions of the Sun at the equinoxes and solstices. \cite[Norris et al (2009)]{Norrisc} have confirmed these alignments and have shown that the straight sides of the circle also indicate the solstices (Fig.\,\ref{wurdi}).
   
      However, a sceptic might still raise some doubts. First, the outliers are only accurate to a few degrees - could these alignments have occurred by chance? Second, although the stones of the circle are large and immovable, the outliers are small and could have been moved. Third, besides the outliers indicating the solstices and equinox, there is an additional outlier whose significance is unclear. The best way to address these doubts would be to find another site with astronomical alignments.  Other stone arrangements in Victoria also indicate the cardinal points, from which we may conclude that the local Aboriginal people knew these directions with some precision, presumably by observing celestial bodies. But are there other sites which point to the position of the solstices? The search continues.
   
\section{Conclusion }
There is a growing body of evidence that traditional Aboriginal people were deeply fascinated by the sky, and the motion of the bodies within it, and had a far richer and deeper knowledge of the sky than is usually appreciated. However, while the evidence for actual measurements or records is suggestive, it remains unproven, although the clues are sufficiently tantalising to fuel the search for more evidence.

\section{Acknowledgement}
This project is dedicated to the hundreds of thousands of Indigenous Australians who lost their lives after the British occupation of Australia in 1788. We especially thank the elders and people of the Yolngu community at Yirrkala, NT, for their help and hospitality, and for their permission to discuss aspects of their culture in our publications. We are also grateful to the other Indigenous groups who have welcomed us onto their land. We also thank Barnaby Norris and Cilla Norris, who helped with much of the research described here, and our colleagues Hugh Cairns, John Clegg, Paul Curnow, Kristina Everett, and John Morieson for their help and collaboration.

\end{document}